\documentclass[iop]{emulateapj-rtx4} % read by: ACROREAD
\shortauthors{Sekanina}
\shorttitle{Nongravitational Accelerations in Long-Period Comets}
\slugcomment{Version \today }

\newcommand{\gapeq}{$\;$\raisebox{0.3ex}{$>$}\hspace{-0.28cm}\raisebox{-0.75ex}{$\sim$}$\;$}

\begin{document}
\title{ON THE DETECTION AND DISTRIBUTION OF NONGRAVITATIONAL ACCELERATIONS\\IN THE MOTIONS OF
       LONG-PERIOD COMETS}
\author{Zdenek Sekanina}
\affil{Jet Propulsion Laboratory, California Institute of Technology,
  4800 Oak Grove Drive, Pasadena, CA 91109, U.S.A.}
\email{Zdenek.Sekanina@jpl.nasa.gov.}

\begin{abstract} % maximum length = 1920 characters; estimate = 1933 (1919)
Even though the orbital motions of most long-period comets are found to comply with the gravitational law, a rapidly increasing minority of these objects is found to display detectable outgassing-driven deviations.  The systematic research of nongravitational effects in the motions of long-period comets began with Hamid \& Whipple's work aimed to support Whipple's icy-conglomerate comet model in the 1950s, profited from fundamentally new results for fragments of the split comets in the 1970s, and has been expanding ever since the development of the Style~II nongravitational orbit-determination model by Marsden et al.\ (1973).  It has also thrived thanks to a dramatic improvement in the quality of astrometric observations in the past decades.  A set of 78~long-period comets with derived nongravitational accelerations and perihelion distances smaller than 2~AU has been assembled and its cumulative distribution shown to mirror the distribution of nuclear dimensions, once the bias toward bright comets and comets with large nongravitational effects is accounted for.  As a rule, long-period comets with detectable deviations from the gravitational law appear to have small, subkilometer-sized nuclei.  Comet C/1995 O1 Hale-Bopp and perhaps some others remain a challenge.
\end{abstract}
\section{Introduction} %%% Section 1
It is well-known that the motions of most long-period comets show no discernible nongravitational effects.  This behavior differs fundamentally from that of the short-period comets, in whose motions such effects are easily perceptible, once an orbital solution links a large enough number of returns to perihelion.  The disparity between the two groups of comets is a consequence of strongly unequal sensitivity of the individual orbital elements to the momentum transferred to the orbital motion by sublimating ices driving the nongravitational acceleration.  Most responsive to such perturbations is the orbital period, which can be determined with extremely high accuracy for comets that return to perihelion every decade or so, whereas the single-apparition comets are greatly disadvantaged.  Both groups are on a par only when it comes to the split comets, as discussed in Section~3.

Pioneering work on the problem of detecting nongravitational motions of long-period comets was carried out by Hamid \& Whipple (1953) in a search for supporting evidence of Whipple's (1950, 1951) icy conglomerate model.  Their effort was based on a simple but ingenious idea to skillfully amend published gravitational orbital solutions to allow the determination of a nongravitational parameter, as described in Section~2.   

\begin{table*}[t]
\vspace{-4.18cm}
\hspace{0.56cm}
\centerline{
\scalebox{1}{
\includegraphics{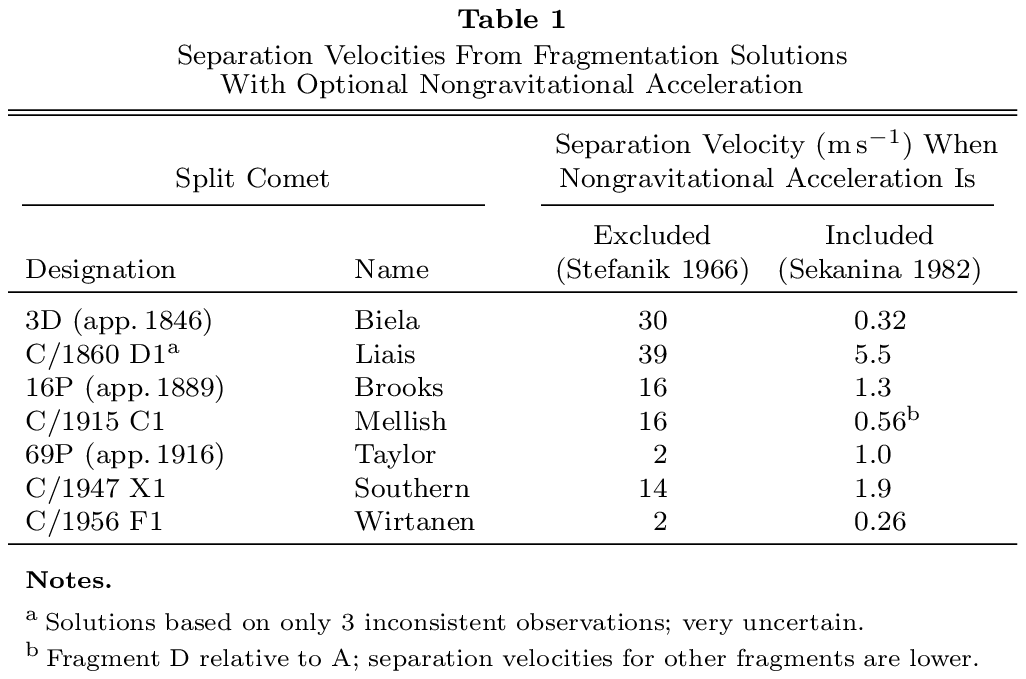}}}
\vspace{-18cm}
\end{table*}

A sophisticated orbit-determination tool and high-quality astrometry were needed to provide compelling evidence for nongravitational effects in the orbital motions of individual long-period comets.  Work on the first issue was successfully pursued by Marsden (1969), who formulated a new orbital code that incorporated the nongravitational terms directly into the equations of motion, using an RTN\footnote{Radial, transverse, normal.} right-handed orthogonal coordinate system, referred to the comet's orbital plane, with the R axis in the anti-solar direction and the N axis toward the north orbital pole.  The task was completed by Marsden et al.\,(1973),
who equipped the computer code with a realistic nongravitational law, tied to the sublimation rate of water ice.  This was the so-called Style II nongravitational model that has since gained universal recognition, becoming the most widely used orbit determination technique for comets worldwide.  I return to further particulars of this model in Section~4.

Given that the averaged vector of the momentum exerted on a cometary nucleus by the sublimating ices points essentially in the direction of the Sun, the dominant component of the nongravitational acceleration points away from the Sun (thus being technically a deceleration in reference to the Sun's gravitational attraction).  Measured by a {\it parameter ${\cal A}$\/}, the magnitude of the effect at 1~AU from the Sun and expressed usually in units of 10$^{-8}$\,AU~day$^{-2}$, it is this {\it radial component of the nongravitational effect on the motions of long-period comets\/} that this study is dealing with in detail.

\section{Hamid \& Whipple's Pioneering Investigation} %%% Section 2
In Hamid \& Whipple's (1953) basic rationale, the outgassing-driven force makes the icy conglomerate comet move in a field of an effectively reduced gravitational attraction of the Sun, which diminishes the Gaussian gravitational constant $k$ by a small amount, \mbox{$\Delta k < 0$}.  If the force is assumed to be approximately proportional to the insolation rate, varying as an inverse square power law of heliocentric distance, $\Delta k$ should for a given comet remain essentially constant.  As long as the gravitational orbital solutions published by the original authors incorporated details of the computations including the normal equations to correct the orbital elements, Hamid \& Whipple amended the equations with a term for $\Delta k/k$ and solved the normal equations for seven, instead of the usual six, orbital elements.  They tabulated these results for a total of 64~long-period comets with perihelion distances not exceeding 1.5~AU and perihelion times between 1850 and 1932.  For 56~percent of the comets the nominal sign of $\Delta k/k$ was negative (and therefore potentially meaningful), but for more than 60~percent the relative error of $\Delta k/k$ exceeded or about equaled $\pm$100~percent, so for these objects the nongravitational effects remained clearly undetected.  Among the 10~individual entries with the lowest absolute errors, not exceeding \mbox{$\pm 1.2 \!\times\! 10^{-5}$}, eight were nominally negative.  The weighted mean came out to be \mbox{$\langle \Delta k/k \rangle = (-0.53 \!\pm\! 0.10) \!\times\! 10^{-5}$}, the half of the set of 64 with perihelion distances smaller than the median (0.72~AU) accounting for nearly the entire effect.  Interestingly, however, the two Kreutz sungrazers in the set showed no measurable effect whatsoever.  Given that in the relevant range of heliocentric distances the differences between the sublimation laws used by Hamid \& Whipple on the one hand and by Marsden et al.\,(1973) on the other hand are relatively small, the dimensionless $\Delta k/k$ effect can be converted to the nongravitational parameter $\cal A$, introduced in Section~1, by
%
% \begin{table*}
% \vspace{-4cm}
% \hspace{0cm}
% \centerline{
% \scalebox{1}{
% \includegraphics{t1_ng.ps}}}
% \vspace{-17cm}
% \end{table*}
%
\begin{equation}
{\cal A} = -29.6 \!\times\! 10^{-5} \frac{\Delta k^2}{k^2} = -59.2 \!\times\! 10^{-5} \frac{\Delta k}{k} \;\;{\rm AU}\;{\rm day}^{-2}, %%% (1)
\end{equation}
so that Hamid \& Whipple's averaged $\Delta k/k$ is equivalent to
\begin{equation}
\langle {\cal A} \rangle = (+0.31 \!\pm\! 0.06) \!\times\! 10^{-8} \, {\rm AU} \; {\rm day}^{-2}, %%% (2)
\end{equation}
to be compared with the pertinent results from the split comets and, especially, from the modern nongravitational solutions below.  

\section{Data from the Split Comets} %%% Section 3
Traditionally, the gradually increasing distance between fragments used to be attributed to a separation velocity.  Unfortunately,  the fit to the fragments' relative motions over longer periods of time was often rather unsatisfactory and the separation velocity came out to be much too high.  More than 40~years ago I proposed a fundamentally different model, which postulated that the growing separation is determined primarily by the difference between the outgassing-driven nongravitational accelerations that the fragments have been subjected to since the time of fragmentation, the separation velocity usually contributing only a small fraction of the effect (Sekanina 1977, 1978).  On a set of seven split comets, Table~1 exemplifies the influence that the introduction of a differential acceleration has on the separation velocity by comparing Stefanik's (1966) results (which ignore the acceleration) with Sekanina's (1982) results (which include the acceleration).

The differential nongravitational accelerations for fragments of 21~split comets (Sekanina 1982) can be converted to the parameter ${\cal A}$ only approximately, because this model --- like Hamid \& Whipple (1953) in their investigation --- applies a nongravitational law that varies as an inverse square of heliocentric distance.  The lowest well-determined acceleration (for fragment B of C/1947~X1 or fragment D of C/1975~V1 is
equivalent to
\begin{equation}
{\cal A} = 1.4 \!\times\! 10^{-8} \; {\rm AU} \; {\rm day}^{-2}, %%% (3)
\end{equation}
with an error of less than $\pm$5~percent.  Since this is a differential value, the (unknown) value for the main fragment (presumably much smaller than the companion's) should still be added.  The peak acceleration (for fragment C of C/1899~E1) was exactly two orders of magnitude higher, with an uncertainty of $\pm$35~percent.  An extremely low acceleration, \mbox{${\cal A} = 0.2 \!\times\! 10^{-8}$\,AU day$^{-2}$}, was obtained for fragment B of the sungrazer C/1965~S1 on the assumption of a zero contribution from the separation velocity, but because of the nature of the orbit, the effects of the separation velocity and the nongravitational acceleration could not be discriminated from each other, so that the value is unlikely to be meaningful.

\section{Style II Nongravitational Model and Its Derivatives} %%% Section 4
Marsden et al.'s (1973) Style II model for the orbital motion of a comet is characterized by a nongravitational law, which is in compliance with a particular regime of water ice sublimation from a~spherical{\vspace{-0.04cm}} nucleus.  At a heliocentric distance $r$, the sublimation rate $\dot{Z}$ from a unit surface element on the nucleus is determined by a steady-state condition that establishes an equilibrium temperature $T$ at which the absorbed fraction of the incident solar radiation is spent in part on the ice sublimation, in part on the black-body reradiation (the heat transfer into subsurface layers being neglected), namely,  
\begin{equation}
\zeta (1 \!-\!A_{\rm B}) \frac{S_0}{r^2} = \mu \dot{Z}(T) \Lambda + \epsilon \sigma T^4, %%% (4)
\end{equation}
where \mbox{$\zeta \leq 1$} is a coefficient that describes the effect of a deviation from normal incidence of the direction of the solar flux received at the surface element (equaling $\cos \theta$ when the Sun's angle with the normal
to the surface is $\theta$), $A_{\rm B}$ is the Bond albedo of the surface, $S_0$ the solar constant, $\mu$ the mass of a water molecule, $\Lambda$ the latent heat of sublimation for water ice, $\epsilon$ the emissivity of the surface, and $\sigma$ the Stefan-Boltzmann constant.{\vspace{-0.05cm}}  In this equation, the sublimation rate $\dot{Z}$ is related to the sublimation pressure $p$ and temperature by
\begin{equation}
\dot{Z} = \frac{p}{\sqrt{2 \pi \mu k_{\rm B} T}},  %%% (5)
\end{equation}
where $k_{\rm B}$ is the Boltzmann constant.  The dependence of the sublimation pressure on the equilibrium temperature is in the first approximation expressed by an Antoine-type equation.

As \mbox{$r \rightarrow 0$}, the first term on the right-hand side of Equation~(4) dominates and{\vspace{-0.03cm}} the sublimation rate varies as essentially $r^{-2}$ at a nearly constant temperature; as \mbox{$r \rightarrow \infty$}, the{\vspace{-0.07cm}} second term dominates and the temperature varies as essentially $r^{-\frac{1}{2}}$.  At any given heliocentric distance, different surface elements have of course different temperatures and different sublimation rates.  The sublimation law employed in the Style~II model is derived with \mbox{$\zeta = \frac{1}{4}$}, which acknowledges that the cross-sectional area of the nucleus exposed to the Sun' radiation is one quarter of the total surface area.   This rather brutal approximation is often referred to as a rapidly-rotating case.

By virtue of its being a function of heliocentric distance, the sublimation law is symmetric with respect to perihelion.  To apply it as a nongravitational law in orbital calculations, it was necessary to find an empirical curve that closely approximates the values of the sublimation rate over as wide a range of heliocentric distances.  Marsden et al.\ (1973) settled on a formula
\begin{equation}
g(r) = \alpha \!\left( \!\frac{r}{r_0} \!\right)^{\!-m} \!\left[ 1 \!+\!
\left( \!\frac{r}{r_0}\!\right)^{\!n} \right]^{\!-k} \!\!, %%% (6)
\end{equation}
where $m$, $n$, $k$, and $r_0$ are constants and $\alpha$ is a normalization factor, so that \mbox{$g(1 {\rm AU}) = 1$}.  It can be written in terms of the four constants and then eliminated from the expression (6) to give
\begin{equation}
g(r) = r^{-m} \!\left( \! \frac{1 \!+\! r_0^n}{r^n \!+\! r_0^n} \!
  \right)^{\!\!k}. %%% (7)
\end{equation}
In order to fit the sublimation rates for the Style~II nongravitational model, the constants equal: \mbox{$m = 2.15$}, \mbox{$n = 5.093$}, \mbox{$nk = 23.5$}, and \mbox{$r_0 = 2.808$ AU}; $r_0$ is a scaling distance.

The function (6) or (7) has a remarkable property. Fitting the sublimation curve of a material other than water ice, the three
exponents do not change dramatically; in particular, \mbox{$m \approx 2$},
\mbox{$nk \approx 20$}, and \mbox{$n \approx k$}.  By contrast, the scaling distance varies considerably with the degree of volatility of the material, depending strongly on the heat of sublimation,
\begin{equation}
r_0 \sim \Lambda^{-2}.  %%% (8)
\end{equation}
For the sublimation rates from the subsolar point the value of $r_0$ is twice as large as it is for the rapidly-rotating case; the rate decreases with $r$ much less steeply.

There have been numerous variations of the Style~II model.
For some comets, including the interstellar object C/2019~Q4 (Borisov), Nakano \& Kobayashi (2020) have been employing essentially the Style~II model, but with a nongravitational law proposed by Yabushita (1996) for highly volatile ices, such as CO.  The reason was apparently the fact that the essentially inverse square power law offered for these objects a somewhat better fit than the standard nongravitational law, whose slope is much steeper already near 2~AU from the Sun.

With my collaborators, I used the above mentioned property of the scaling distance on several occasions in the past, extensively in an investigation of
several dwarf Kreutz sungrazers observed with the C2 coronagraph on board the
Solar and Heliospheric Observatory (Sekanina \& Kracht 2015).  We found that for some of them the nongravitaional laws were extremely steep, with scaling distances as short as 0.012~AU, close to the expected sublimation curve for olivine.  On the other hand, a similar exercise for the extensively observed comet C/1995~O1 (Sekanina \& Kracht 2017) resulted in only a slight deviation of the scaling distance from the standard value.

Yeomans \& Chodas (1989) applied an asymmetric nongravitational law to several short-period comets.  The model was based on Sekanina's (1988) idea proposed in his work on Encke's comet:\ the sublimation rate for any time $t$ is taken not at
a heliocentric distance $r(t)$, but at $r(t\!-\!\Delta t)$.  The time interval $\Delta t$ is an additional nongravitational parameter.  When $\Delta t$ is positive, the nongravitational curve peaks after perihelion; when it is negative, the peak occurs before perihelion.  Yeomans \& Chodas successfully applied this model to several short-period comets, finding that the radial nongravitational acceleration always dominated.  The model was not employed for any long-period comet.

Still more complex modifications of the Style~II model were developed and applied by Sitarski (1994a, 1994b), who was able to link the motion of 22P/Kopff in 13 successive apparitions over a period of 84~years by accounting for the nongravitational effects triggered by the comet's precession, as investigated by Sekanina (1984).  Other, earlier attempts to modify the Style~II model were less successful than Sitarski, Yeomans, and Chodas.

As is apparent from these examples, the Style~II model and its many variations were, especially in the first decades after the model's formulation, applied predominantly to the motions of short-period comets.  This is understandable because too few long-period comets with the motions affected by the nongravitational acceleration were known to have good statistics.  Indeed, the set presented in the next section shows that more than 50~percent of the long-period comets with nongravitational effects come from the years starting in 2000 and 70~percent from the years starting{\nopagebreak} in 1990.  It is unquestionably the dramatically improving quality of astrometric data over the past two to three decades that is reflected in these numbers.

\begin{figure}[t]
\vspace{-2.1cm}
\hspace{0.8cm}
\centerline{
\scalebox{0.63}{
\includegraphics{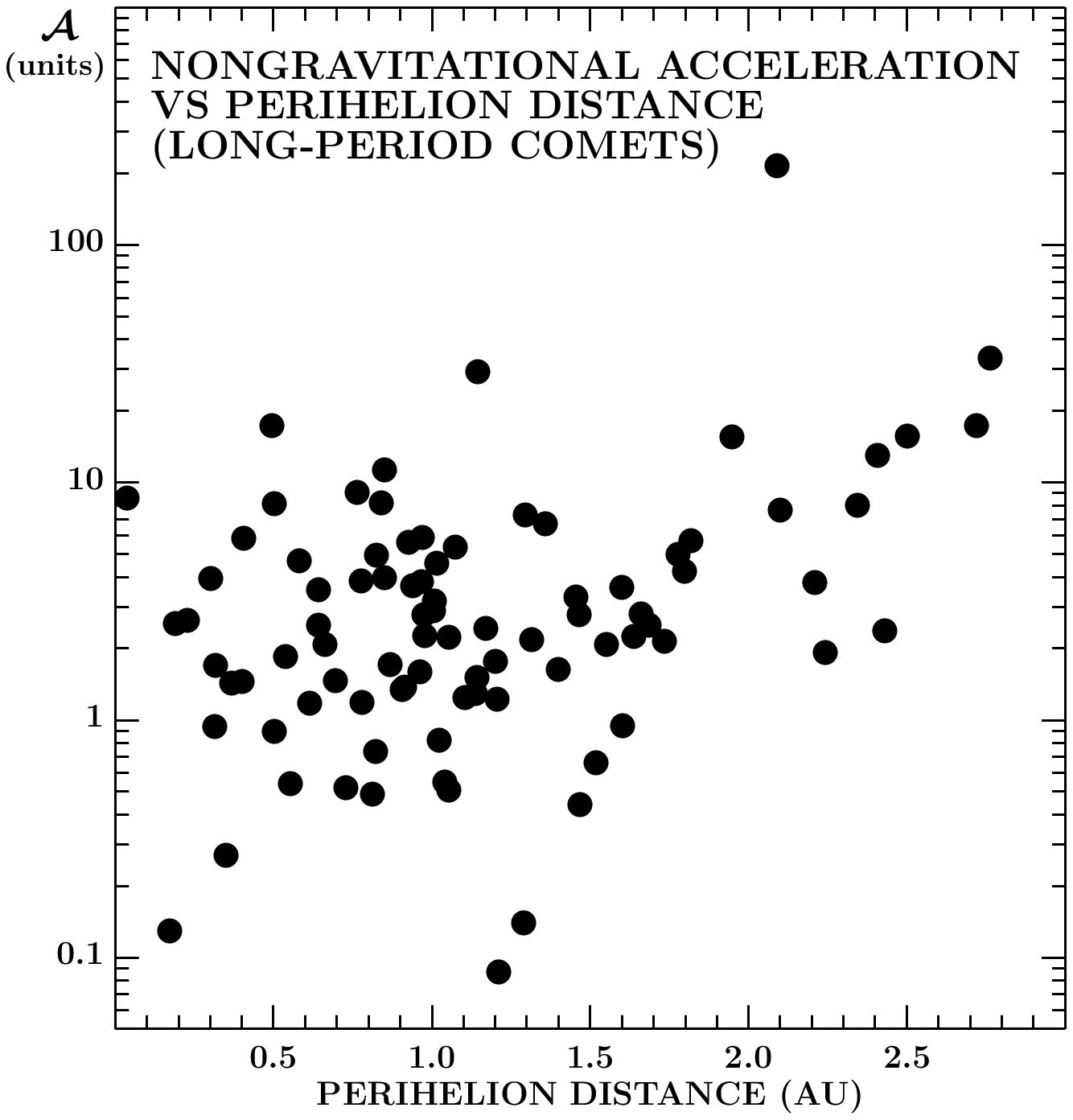}}}
\vspace{-7.95cm}
\caption{Plot of the nongravitational{\vspace{-0.045cm}} parameter $\cal A$ (in units of
 10$^{-8}$\,AU~day$^{-2}$) for 88~long-period comets against the perihelion
 distance.  Note that for perihelion distances larger than 2~AU, non\-gravitational
 parameters smaller than 1~unit are missing.{\vspace{0.5cm}}}
\end{figure}

\section{Representative Set of Nongravitational Orbits of Long-Period Comets}
For the purpose of examination of the distribution of nongravitayional accelerations affecting the motions of long-period comets, I consulted four major sources as follows: \\

(1) the online {\it Jet Propulsion Laboratory's Small Body Database Search Engine\/},\footnote{See {\tt
https://ssd.jpl.nasa.gov/sbdb.query.cgi}.} which contributed 47~entries; \\[-0.1cm]

(2) the Warsaw {\it Catalogue of Cometary Orbits and Their Dynamical Evolution\/} (CODE); 45~contributed entries are from Kr\'olikowska (2014) and Kr\'olikowska et al.\ (2014);\footnote{At the time of completion of this part of the work, the final version of the Catalogue (Kr\'olikowska 2020, Kr\'olikowska \& Dybczy\'nski 2020) was not yet available, but the 2014 papers appear to contain the majority of the data.} \\[-0.1cm]
 
(3) the online notes by Nakano,\footnote{See {\tt http://www.oaa.gr.jp/$\sim$oaacs.nk.htm}.} listing the orbital elements of well observed comets, also contributing 45 entries; and \\[-0.1cm]

(4) {\it Catalogue of Cometary Orbits\/} by Marsden \& Williams (2008), which has contributed 28~entries. \\

As noted, only the orbital solutions based on the Style II nongravitational law have been considered.  A combined list of the $\cal A$ parameters from these sources includes information on 100~long-period comets with perihelion distances smaller than 3~AU.  The comets with perihelia farther from the Sun have been eliminated from the data set because their $\cal A$ parameters were systematically much too high, the result of dominance of very volatile ices.  Of the 100~comets, 12 that had a negative or poorly determined $\cal A$, or the transverse component higher than or comparable to the radial component of the acceleration, were removed, thereby reducing the working set to 88~entries.  These data, plotted against the perihelion distance in Figure~1, showed that excessively high values of $\cal A$ prevailed down to 2~AU from the Sun, necessitating removal of 10~comets with perihelia exceeding 2~AU.  The final set consisting of 78~comets is listed chronologically in Table~2.  The last column identifies the source by an abbreviation:\ J for the JPL database; W for the Warsaw catalogue; N for the Nakano notes (with the number); and C for the Marsden--Williams catalogue.  Interestingly, values of $\cal A$ by all four data sources have been available for only one comet; for 11~comets by three sources; and for 26~comets by two sources.  For 37~comets just a single value of $\cal A$ has been available.  In most instances the agreement among the various determinations is rather satisfactory, well within one unit of each other.  In a few cases, however, the results are less satisfactory; for example, for C/2002~T7 the values of $\cal A$ (in units of 10$^{-8}$\,AU~day$^{-2}$) by JPL (+1.21) and Nakano (+1.14) agree very well, but the values in the CODE catalogue (+0.38) and the Marsden--Williams catalogue ($-$0.19) differ.  In the multiply derived values of $\cal A$, the value from the source prioritized in the order from (1) through (4) above was preferred, but when they closely agreed, they were averaged.

\section{Distribution of Nongravitational Accelerations in the Motions of Long-Period Comets and Its Modeling}

A log-log cumulative distribution of the $\cal A$ parameters for the 78 long-period comets is plotted in Figure~2.  It suggests that the number of comets with nongravitational effects exceeding $\cal A$ varies approximately as ${\cal A}^{-1.6}$ for
\mbox{${\cal A} {\mbox{\gapeq}} 4 \times \!
10^{-8}$\,AU day$^{-2}$}, but progressively less steeply at lower
accelerations, presumably because of growing difficulties to detect
minor nongravitational effects.

In an effort to model the distribution in Figure 2, I recall that the
conservation of momentum law predicts a general relationship for the
nongravitavitational acceleration $\cal A$:
\begin{equation}
{\cal A} = \epsilon_0 \frac{v_{\rm subl} \dot{\cal M}_{\rm subl}}{\cal
 M},
\end{equation}
where $\cal M$ is the mass of the nucleus, $\dot{\cal M}_{\rm subl}$
is its mass rate of gas sublimated away per unit time at a velocity of
$v_{\rm subl}$, and $\epsilon_0$ is an efficiency factor, including,
for example, the fraction of the surface active, the collimation of
the outflow, etc.  For a spherical nucleus of diameter $D$ one has of
{\vspace{-0.06cm}}course \mbox{${\cal M} \sim D^3$}.  Furthermore,
I assume that \mbox{$\dot{\cal M}_{\rm subl} \sim D^2$} and that
$v_{\rm subl}$, approximated by the speed of sound at the nuclear
surface, is independent of $D$.  Equation~(9) then becomes
\begin{table*}
\vspace{-4.18cm}
\hspace{0.52cm}
\centerline{
\scalebox{1}{
\includegraphics{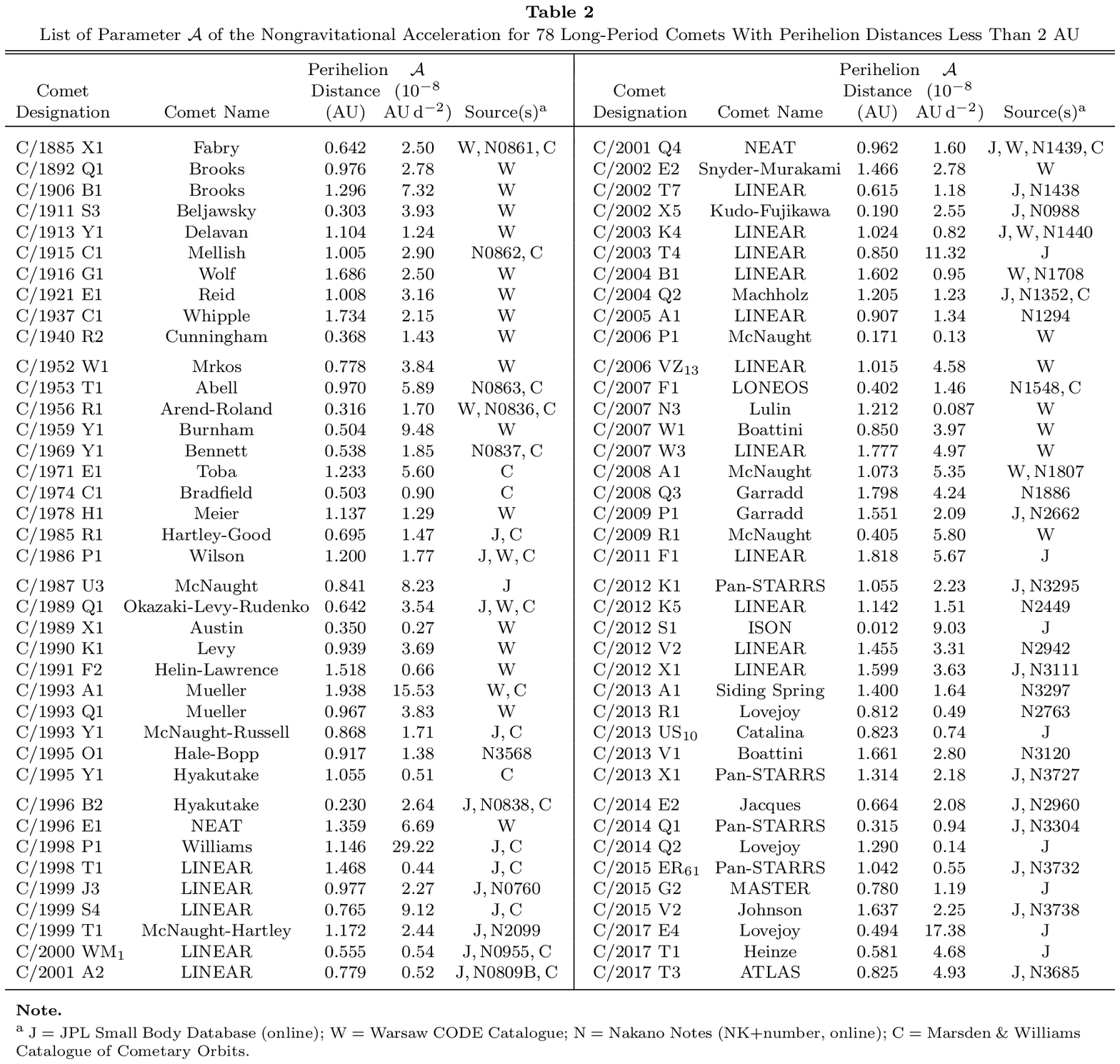}}}
\vspace{-8cm}
\end{table*}
\begin{equation}
{\cal A} = \frac{\epsilon}{D},
\end{equation}
where $\epsilon$ is a constant; for water ice outgassing, its value
is about 0.5\,$\times 10^{-8}$ when $\cal A$ is in AU~day$^{-2}$ and
$D$ in km.

Formula (10) indicates that the distribution of $\cal A$ is closely
related to the nuclear size distribution.  Lamy et al.\ (2004) address
this latter issue and show that the number of comets, $dN(D)$, with
diameters between $D$ and $D \!+\! dD$ can be expressed by a power law,
\begin{equation}
dN(D) = c D^{-\nu} dD
\end{equation}
where different studies, based on analysis of the luminosity function, yield
for $\nu$ values\footnote{Lamy et al.\ (2004) presented their results in
terms of the cumulative distribution; the power laws they list are here
converted to the respective differential distributions.} in a general range
from 2.6 to 3.7.

In order to fit the observed distribution of $\cal A$, I account for
two major selection effects:\ (a)~the data set is strongly biased toward
bright comets, because the nongravitational terms can be incorporated
into the equations of motion only for comets observed over long periods
of time; and (b)~the data set is also biased toward comets with large
nongravitational effects, because a gravitational solution otherwise
fits astrometric observations adequately and no nongravitational solution
is even attempted.  The bias (a) requires that the distribution (11) be
modified by incorporating a factor $p_1(\Im)$, where $\Im$ is the intrinsic
brightness of the comets, the bias (b) necessitates the inclusion of a
factor $p_2({\cal A})$, which converges to
\begin{eqnarray}
\lim_{{\cal A} \rightarrow 0} p_2(\cal A) & = & 0, \nonumber \\
\lim_{{\cal A} \rightarrow \infty} p_2(\cal A) & = & 1.
\end{eqnarray}
\begin{figure}
\vspace{-3.34cm}
\hspace{1.85cm}
\centerline{
\scalebox{0.78}{
\includegraphics{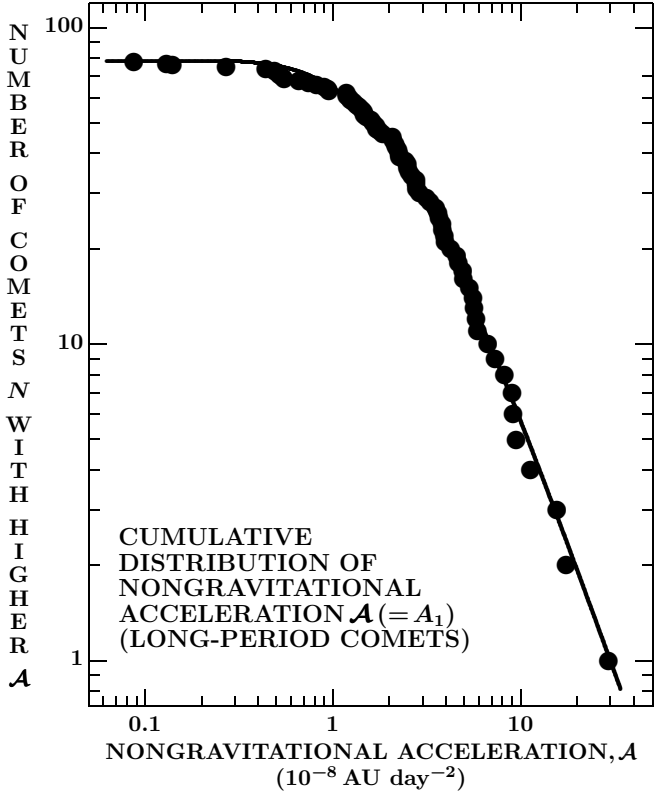}}}
\vspace{-9.47cm}
\caption{Cumulative distribution of 78 long-period comets with perihelion
distances smaller than 2~AU. whose nongravitational accelerations are higher
than $\cal A$.  The fit is described by the expression (16), the parameters
are listed by Eauation~(18).{\vspace{0.45cm}}}
\end{figure}

\noindent
The distribution (11) is now modified thus
\begin{equation}
dN(D,\Im,{\cal A}) = c D^{-\nu} p_1(\Im) \, p_2({\cal A}) \, dD.
\end{equation}
In the following I adopt $p_1(\Im) = b_0 \Im^\beta$ ($\beta > 0$);
since, in general, \mbox{$\Im \!\sim\! D^2$}, one has
\begin{equation}
p_1(\Im) = b D^{2\beta}.
\end{equation}
For $p_2({\cal A})$ I use an exponential function that satisfies the
conditions (12), namely
\begin{equation}
p_2({\cal A}) = \exp \left( \! - \frac{a}{\cal A} \right).
\end{equation}
Writing next Equation (13) in terms of the expressions in (14) and (15),
substituting $\cal A$ for $D$ from (10), and integrating over all $\cal A$
greater than ${\cal A}_0$, the cumulative distribution is given by
\begin{eqnarray}
N({\cal A}_0) & = & C_0 \!\int_{{\cal A}_0}^{\infty} \!\! {\cal A}^{\nu - 2(\beta
 + 1)} \exp\!\left( \!-\frac{a}{\cal A} \right) d{\cal A} \nonumber \\
 & = & C \!\int_{0}^{(a\!/\!{\cal A}_0)} \!\! u^{2\beta - \nu} e^{-u} du \nonumber \\[0.1cm]
 & = & C\,{\mbox{\boldmath $\gamma \:\!\!\!\!\! \gamma$}} \!\left(\! 2\beta
 \!+\! 1 \!-\! \nu,\frac{a}{{\cal A}_0} \! \right),
\end{eqnarray}
where \mbox{$C_0 = b\:\!c\:\! \epsilon^{2\beta \!+\! 1 \!-\! \nu}$},
\begin{equation}
C = b\:\!c\:\! (\epsilon/a)^{2\beta \!+\! 1 \!-\! \nu} = \frac{N_0}{\Gamma(2\beta
 \!+\! 1 \!-\! \nu)},
\end{equation}
\mbox{$N_0 = \lim_{{\cal A}_0 \rightarrow 0}\, N({\cal A}_0) = 78$}, $\Gamma(x)$
is the Gamma function of argument $x$, and \mbox{\boldmath
$\gamma\!\!\!\!\!\:\gamma$}$(x,y)$ is the incomplete gamma function of the first
kind.

As shown in Figure 2, an excellent fit has been achieved with the
following constants:
\begin{eqnarray}
2\beta \!+\! 1 \!-\! \nu & = & 1.7, \nonumber \\
a & = & 3.1 \!\times\! 10^{-8} \,{\rm AU \,\, day}^{-2}, \nonumber \\
C & = & 85.84, \nonumber \\
b\:\!c & \simeq & 1900;
\end{eqnarray}
$b$ and $c$ cannot be separated from one another, unless one makes
some very problematic assumptions about detection of the nongravitational
acceleration in the orbital motions of the category of comets under
consideration; the values of the two constants are of no particular
interest to this investigation.

Adopting with Lamy et al.\,(2004) that \mbox{$2.6 \!\leq\! \nu
\!\leq\! 3.7$}, one finds that \mbox{$1.65 \!\leq\! \beta \!\leq\! 2.2$},
so that the chance of detecting the nongravitational acceleration in the
orbital motions of long-period comets varies approximately as the
square of their intrinsic brightness, assuming the geocentric distances
average out.

The distribution (16) can also be used to estimate a median
nongravitational acceleration $\langle {\cal A} \rangle$,
which is defined by a condition:
\begin{equation}
\gamma \hspace{-0.216cm} \gamma \left( \! 2\beta\!+\!1\!-\!\nu,
 \frac{a}{\langle {\cal A} \rangle} \! \right) = {\textstyle
 \frac{1}{2}} \Gamma(2\beta \!+\!1\!-\!\nu).
\end{equation}
The solution to this condition is \mbox{$a/\langle {\cal A} \rangle
= 1.38076$} and \mbox{$\langle {\cal A} \rangle = 2.2 \times \!
10^{-8}$\,AU day$^{-2}$}.  However, inspection of catalogs of
cometary orbits shows that the number of long-period comets
with \mbox{$q < 2$ AU} observed since 1885 (excluding comets
discovered with SOHO and other space observatories) is
approximately 230, so only for 34~percent of them the
nongravitational acceleration was determined.

The cumulative distribution is next expanded by incorporating
the remaining 66~percent, or 152~long-period comets with
only gravitational orbits on the assumptions that (i)~the
distribution of the missing nongravitational parameters
would satisfy the expressions (10) through (16) with the values
for $\beta$ and $\nu$ unchanged; and (ii)~for none of the
additional 152~comets could the value of ${\cal A}$ exceed
the maximum value ${\cal A}_{\rm max}$ plotted in Figure~5.
For \mbox{$N({\cal A}_{\rm max}) = 1$} the distributon
function (16) yields:
\begin{equation}
\gamma \hspace{-0.216cm} \gamma \left( \! 2\beta \!+\! 1 \!-\! \nu,
 \frac{a}{{\cal A}_{\rm max}} \! \right) = \frac{\Gamma(2\beta
 \!+\! 1 \!-\! \nu)}{N_0} = 0.00395,
\end{equation}
where one now has \mbox{$N_0 = 230$}.  With \mbox{$2\beta \!+\! 1
\!-\! \nu = 1.7$} the condition (20) implies that for the expanded
distribution \mbox{$a/{\cal A}_{\rm max} = 0.05376$} and
\mbox{$a = 1.57\,\times\:\!$10$^{-8}$\,AU~day$^{-2}$}.
From Equation~(19) one finds an improved value for the median
nongravitational acceleration
\begin{equation}
\langle {\cal A} \rangle = 1.14 \times \:\!\! 10^{-8}\,{\rm AU
 \,\, day}^{-2}.
\end{equation}
From Equation (10) one finds a median nuclear diameter of a little less than 0.5~km.

\section{Conclusions}
Investigations of nongravitational effects in the motions of long-period comets have a very short history.  They were initiated by the work of Hamid \& Whipple (1953), have benefited from studies of the split comets, and expanded greatly following the formulation of the Style~II nongravitational model by Marsden et al.\ (1973) and thanks to the greatly improved quality of astrometric observations of comets in the past decades that allows detection of minor systematic trends in the residuals from gravitational solutions.

While the nongravitational accelerations vary greatly from comet to comet, the component in the antisolar direction practically always dominates.  Typical values of the parameter $\cal A$ are on the order of 10$^{-8}$\,AU~day$^{-2}$, suggesting effects driven by sublimation of water ice from subkilometer-sized cometary nuclei.  Modeling of the cumulative distribution of the nongravitational parameter $\cal A$ of 78~long-period comets with perihelion distances smaller than 2~AU shows that it reflects their nuclear-size distribution, once the bias toward bright comets and comets with large nongravitational effects is accounted for.  It is concluded that, as a rule, long-period comets whose motions display perceptible deviations from the gravitational law have nuclei only hundreds of meters or less across.  However, some comets, and notably C/1995~O1 Hale-Bopp, remain a challenge.\\

This research was carried out at the Jet Propulsion Laboratory, California Institute of Tachnology, under contract with the National Aeronautics and Space Administration.

\begin{center}
{\footnotesize
REFERENCES}
\end{center}
\begin{description}
{\footnotesize
% Guigay, G. 1955, J. Obs., 38, 189
%
\item[\hspace{-0.3cm}]
Hamid, S. E., \& Whipple, F. L. 1953, AJ, 58, 100
\\[-0.57cm]
\item[\hspace{-0.3cm}]
Kr\'olikowska, M. 2014, A\&A, 567A, 126
\\[-0.57cm]
\item[\hspace{-0.3cm}]
Kr\'olikowska, M. 2020, A\&A, 633A, 80
\\[-0.57cm]
\item[\hspace{-0.3cm}]
Kr\'olikowska, M., \& Dybczy\'nski, P. A. 2020, A\&A, 640A, 97
\\[-0.57cm]
\item[\hspace{-0.3cm}]
Kr\'olikowska, M., Sitarski, G., Pittich, E. M., et al. 2014, A\&A,{\linebreak}
 {\hspace*{-0.6cm}}571A, 63
\\[-0.57cm]
\item[\hspace{-0.3cm}]
Lamy, P. L., Toth, I., Fernandez, Y. R., \& Weaver, H. A. 2004, in{\linebreak}
 {\hspace*{-0.6cm}}Comets II, ed. M. Festou, H. U. Keller, \& H. A. Weaver
 (Tucson,{\linebreak}
 {\hspace*{-0.6cm}}AZ: University of Arizona), 223
\\[-0.57cm]
\item[\hspace{-0.3cm}]
Marsden, B. G. 1969, AJ, 74, 720
\\[-0.57cm]
\item[\hspace{-0.3cm}]
Marsden, B. G., \& Williams, G. V. 2008, Catalogue of Cometary{\linebreak}
 {\hspace*{-0.6cm}}Orbits 2008, 17th ed. Cambridge, MA:\ Minor Planet Center/{\linebreak}
 {\hspace*{-0.6cm}}Central Bureau for Astronomical Telegrams, 195pp
\\[-0.57cm]
\item[\hspace{-0.3cm}]
Marsden, B. G., Sekanina, Z., \& Yeomans, D. K. 1973, AJ,~78,~211
\\[-0.57cm]
\item[\hspace{-0.3cm}]
Nakano, S., \& Kobayashi, T. 2020, Nakano Notes NK 4121, 4161,{\linebreak}
 {\hspace*{-0.6cm}}4167, 4211, 4252, 4253, 4321
\\[-0.57cm]
\item[\hspace{-0.3cm}]
Sekanina, Z. 1977, Icarus, 30, 574
\\[-0.57cm]
\item[\hspace{-0.3cm}]
Sekanina, Z. 1978, Icarus, 33, 173
\\[-0.57cm]
\item[\hspace{-0.3cm}]
Sekanina, Z. 1982, in Comets, ed. L. L. Wilkening (Tucson, AZ:{\linebreak}
 {\hspace*{-0.6cm}}University of Arizona), 251
\\[-0.57cm]
\item[\hspace{-0.3cm}]
Sekanina, Z. 1984, AJ, 89, 1573
\\[-0.57cm]
\item[\hspace{-0.3cm}]
Sekanina, Z. 1988, AJ, 96, 1455
\\[-0.57cm]
\item[\hspace{-0.3cm}]
Sekanina, Z., \& Kracht, R. 2015, ApJ, 801, 135
\\[-0.57cm]
\item[\hspace{-0.3cm}]
Sekanina, Z., \& Kracht, R. 2017, arXiv eprint 1703.00928
\\[-0.57cm]
\item[\hspace{-0.3cm}]
Sitarski, G. 1994a, Acta Astron., 44, 91
\\[-0.57cm]
\item[\hspace{-0.3cm}]
Sitarski, G. 1994b, Acta Astron., 44, 417
\\[-0.57cm]
\item[\hspace{-0.3cm}]
Stefanik, R. P. 1966, M\'em. Soc. Roy. Sci. Li\`ege, 12 (S\'er. 5), 29
\\[-0.57cm]
\item[\hspace{-0.3cm}]
Whipple, F. L. 1950, ApJ, 111, 375
\\[-0.57cm]
\item[\hspace{-0.3cm}]
Whipple, F. L. 1951, ApJ, 113, 464
\\[-0.57cm]
\item[\hspace{-0.3cm}]
Yabushita, S. 1996, Mon. Not. Roy. Astron. Soc., 283, 347
\\[-0.65cm]
\item[\hspace{-0.3cm}]
Yeomans, D. K., \& Chodas, P. W. 1989, AJ, 98, 1083}
\vspace{-0.4cm}
\end{description}
\end{document}